\documentclass[aps,prd,superscriptaddress,twocolumn,showpacs]{revtex4}
\usepackage{graphicx}
\usepackage{epstopdf}
\usepackage{amsmath}
\usepackage{amsfonts}
\usepackage{amssymb}
\usepackage{latexsym}

\begin{document}
\title{A note on nonlinear electrodynamics}
\author{Patricio Gaete} \email{patricio.gaete@usm.cl} 
\affiliation{Departamento de F\'{i}sica and Centro Cient\'{i}fico-Tecnol\'ogico de Valpara\'{i}so-CCTVal, Universidad T\'{e}cnica Federico Santa Mar\'{i}a, Valpara\'{i}so, Chile}
\author{Jos\'{e} A. Helay\"{e}l-Neto}\email{helayel@cbpf.br}
\affiliation{Centro Brasileiro de Pesquisas F\'{i}sicas (CBPF), Rio de Janeiro, RJ, Brasil} 
\date{\today}

\begin{abstract}
We explore the physical consequences of a new nonlinear electrodynamics, for which the electric field of a point-like charge is finite at the origin, as in the well-known Born-Infeld electrodynamics. However, contrary to the latter, in this new electrodynamics the phenomenon of birefringence and dichroism take place in the presence of external magnetic fields. Subsequently we study the interaction energy, within the framework of the gauge-invariant but path-dependent variables formalism. Interestingly enough, the static potential profile contains a linear potential leading to the confinement of static charges. 
\end{abstract}
\maketitle

\section{Introduction}

One of the most startling predictions of Quantum Electrodynamics (QED) is the light-by-light scattering in vacuum and its physical consequences such as vacuum birefringence and
vacuum dichroism \cite{Bamber,Burke,nphoton,Tommasini1,Tommasini2}. 
Very recently, the ATLAS Collaboration has reported on the direct detection of the light-by-light scattering in LHC Pb-Pb collisions with a $4.4$ $\sigma$ level of confidence \cite{Atlas}. Actually, this proposal to look for light-by-light scattering in ultra-peripheral heavy-ion collisions at the LHC was pioneered in Ref. \cite{Enterria}. More recently, inspired by these results was obtained a bound on the $\beta$ parameter in the nonlinear Born-Infeld electrodynamics \cite{Ellis}.

On the other hand, in recent times nonlinear electrodynamics have been the object of intensive investigations  in the context of black hole physics \cite{Fan,Junior}. The interest in studying these electrodynamics is mainly due to the possibility of constructing exact (regular) black hole solutions. In addition, nonlinear electrodynamics has also attracted interest in order to explain the Rindler acceleration as a nonlinear electromagnetic effect \cite{Halilsoy}.

In this perspective we also recall that, in previous works \cite{Nonlinear,Logarithmic,Nonlinear2}, we have studied the physical effects presented by different models of $(3+1)$-D nonlinear Electrodynamics in vacuum. In such a case, it was shown that for Generalized Born-Infeld, and Logarithmic Electrodynamics the field energy of a point-like charge is finite. It should be further noted that Generalized Born-Infeld, Exponential, Logarithmic and Massive Euler-Heisenberg-like Electrodynamics display the vacuum birefringence phenomenon.

Motivated by these observations and given the ongoing experiments related to light-by-light scattering,
it should be interesting to acquire a better understanding what might be the observational signatures presented by vacuum electromagnetic nonlinearities. Hence, our purpose here is to consider a new nonlinear electrodynamics and investigate aspects of birefringence and dichroism, as well as the computation of the static potential along the lines of \cite{Nonlinear,Logarithmic,Nonlinear2}, which is an alternative to the Wilson loop approach.

Our work is organized as follows: in Section II, we describe this new nonlinear electrodynamics and study aspects of birefringence and dichroism. In Section III we calculate the interaction energy for a fermion-antifermion pair. Interestingly enough, the static potential profile contains a linear term, leading to the confinement of static charges. Finally, some concluding remarks are made in Sec. IV. 

In our conventions the signature of the metric is ($+1,-1,-1,-1$).

\section{The model under consideration} 

In this section, we begin our analysis with a brief description of the model under consideration. This would not only provide the theoretical setup for our subsequent work, but also fix the notation. In this case the corresponding model is governed by the Lagrangian density:   
\begin{equation}
{\cal L} =  - {\lambda}^{2} \frac{{\sqrt { - {\cal F}} }}{{\left[ {\sqrt 2 \lambda  + \sqrt { - {\cal F}} } \right]}} , \label{NLM05}
\end{equation}
where ${\cal F} = \frac{1}{4}F_{\mu \nu } F^{\mu \nu }$, and $\tilde F^{\mu \nu }  = \frac{1}{2}\varepsilon ^{\mu \nu \rho \lambda } F_{\rho \lambda }$ is the dual electromagnetic field strength tensor. The $\eta$ constant has ${\left( {mass} \right)^4}$ dimension whereas $\lambda$ has ${\left( {mass} \right)^2}$ dimension in natural units. Let us also mention here that, in a purely electric case, the $\lambda$ constant plays the role of an uniform background electric field as it shall become clear in what follows.

The parameter $\lambda$ must be positive and it is a sort a cut-off for the electric and magnetic fields, $\lambda  \gg |{\bf E}|$ and $|{\bf B}|$. Also, we must stress that our effective model only applies for electromagnetic fields such that
$- \frac{1}{4}{F_{\mu \nu }}{F^{\mu \nu }} = \frac{1}{2}({{\bf E}^2} - {{\bf B}^2}) \ge 0$.

The mass scale, $M$, fixed by the $\lambda$-parameter $(M = \sqrt \lambda)$, characterizes a regime where the non-linearity of the electromagnetic interaction becomes relevant. We can therefore associate ${\lambda ^{ - {\raise0.5ex\hbox{$\scriptstyle 1$}\kern-0.1em/\kern-0.15em\lower0.25ex\hbox{$\scriptstyle 2$}}}}$ to the length scale that appears in brane scenarios, where the particles can be localized in lower-dimensional branes which are separated from each other by a distance in the ${\lambda ^{ - {\raise0.5ex\hbox{$\scriptstyle 1$}\kern-0.1em/\kern-0.15em\lower0.25ex\hbox{$\scriptstyle 2$}}}}$-scale. We believe $\gtrsim 1$ TeV.

Accordingly the field equations read:
\begin{equation}
{\partial _\mu }\left[ {\frac{{\lambda}^{3}}{{2\sqrt 2 }}\frac{{{F^{\mu \nu }}}}{{\sqrt { - {\cal F}} {{\left[ {\sqrt 2 \lambda  + \sqrt { - {\cal F}} } \right]}^2}}}} \right] = 0, \label{NLM15}
\end{equation}
while the Bianchi identity is given by
\begin{equation}
\partial _\mu  \tilde F^{\mu \nu }  = 0. \label{NLM20}
\end{equation}

It should be further noted that Gauss' law reduces to,
\begin{equation}
\nabla  \cdot  (\frac{{\lambda}^{3}}{{\sqrt {{{\bf E}^2} - {{\bf B}^2}} \left[ {2\lambda  + \sqrt {{{\bf E}^2} - {{\bf B}^2}} } \right]^2}}{\bf E} ) = 0, \label{NLM25}
\end{equation}
From equation (\ref{NLM25}) it follows that, for an external point-like charge sitting at the origin, the ${\bf D}$-field lies along the radial direction and is given by ${\bf E} = \frac{Q}{{r^2 }}\hat r$, where $Q = \frac{e}{{4\pi }}$. It is also important to observe that for a point-like charge, e, at the origin, the electrostatic field is given by
\begin{equation}
|{\bf E}| = - 2\lambda  + {\left( {\frac{{\lambda}^{3}}{Q}} \right)^{\frac{1}{2}}}r . \label{NLM35}
\end{equation}

In order to write the dynamical equations into a more compact and convenient form,
we shall introduce the vectors ${\bf D} = {{\partial {\cal L}} \mathord{\left/
 {\vphantom {{\partial L} {\partial {\bf E}}}} \right.
 \kern-\nulldelimiterspace} {\partial {\bf E}}}$ and ${\bf H} =  - {{\partial {\cal L}} \mathord{\left/
 {\vphantom {{\partial L} {\partial {\bf B}}}} \right.
 \kern-\nulldelimiterspace} {\partial {\bf B}}}$, in analogy to the electric displacement and magnetic field strength. We then have 
\begin{equation}
{\bf D} = - \frac{{2{\lambda}^{3}}}{{\sqrt {{{\bf E}^2} - {{\bf B}^2}} {{\left[ {2\lambda  + \sqrt {{{\bf E}^2} - {{\bf B}^2}} } \right]}^2}}}{\bf E} , \label{NLM40}
\end{equation}
and
\begin{equation}
{\bf H} = - \frac{{2{\lambda}^{3}}}{{\sqrt {{{\bf E}^2} - {{\bf B}^2}} \left[ {2\lambda  + \sqrt {{{\bf E}^2} - {{\bf B}^2}} } \right]}}{\bf B} . \label{NLM45}
\end{equation}

With this, we can write the corresponding equations of motion as
\begin{equation}
\nabla  \cdot {\bf D} = 0, \  \  \
\frac{{\partial {\bf D}}}{{\partial t}} - \nabla  \times {\bf H} = 0, \label{NLM50a}
\end{equation}
and
\begin{equation}
\nabla  \cdot {\bf B} = 0, \  \  \
\frac{{\partial {\bf B}}}{{\partial t}} + \nabla  \times {\bf E} = 0. \label{NLM50b}
\end{equation}

It is now important to notice that the complicated field problem can be greatly simplified if the above equations are linearized. As is well-known, this procedure is justified for the description of a weak electromagnetic wave $({\bf E_p}, {\bf B_p})$ propagating in the presence of a strong constant external field $({\bf E_0}, {\bf B_0})$. For computational simplicity our analysis will be developed in the case of a purely magnetic field, that is, ${\bf E_0}=0$. This then implies that  
\begin{equation}
{\bf D} = \Gamma {\bf E}_p, \label{NLM55}
\end{equation}
and
\begin{equation}
{\bf H} = \Gamma \left[ {{{\bf B}_p} - i\frac{\lambda }{{2 \Gamma {{\left( {{\bf B}_0^2} \right)}^{{\raise0.7ex\hbox{$3$} \!\mathord{\left/
 {\vphantom {3 2}}\right.\kern-\nulldelimiterspace}
\!\lower0.7ex\hbox{$2$}}}}}}\left( {{{\bf B}_p} \cdot {{\bf B}_0}} \right){{\bf B}_0}} \right], \label{NLM60} 
\end{equation}
with
\begin{equation}
\Gamma  = \frac{1}{2}\left( {1 + \frac{{i\lambda }}{{\sqrt {{\bf B}_0^2} }}} \right), \label{NLM65} 
\end{equation} 
where we have keep only linear terms in ${\bf E_p}$, ${\bf B_p}$.

\begin{equation}
{\varepsilon _{ij}} = \Gamma {\delta _{ij}} , \label{NLM60a}
\end{equation}
and
\begin{equation}
{\left( {{\mu ^{ - 1}}} \right)_{ij}} = \Gamma \left( {{\delta _{ij}} - \frac{{i\lambda }}{{2 \Gamma {{\left( {{\bf B}_0^2} \right)}^{{\raise0.5ex\hbox{$\scriptstyle 3$}
\kern-0.1em/\kern-0.15em
\lower0.25ex\hbox{$\scriptstyle 2$}}}}}}{B_{0i}}{B_{0j}}} \right). \label{NLM65b}
\end{equation}

Next, without restricting generality we take the $z$ axis as the direction of the magnetic field, ${\bf B_0}  = B_0 {\bf e}_3$, and assuming that the light wave moves along the $x$ axis. We further make a plane wave decomposition for the fields $E_p$ and $B_p$, that is, 
\begin{equation}
{{\bf E_p}}\left( {{\bf x}
,t} \right) = {\bf E}
{e^{ - i\left( {wt - {\bf k} \cdot {\bf x}} \right)}}, \ \ \
{{\bf B_p}}\left( {{\bf x},t} \right) = {\bf B}{e^{ - i\left( {wt - {\bf k} \cdot {\bf x}} \right)}}, \label{NLM70}
\end{equation}
so that the Maxwell equations become
\begin{equation}
\left( {\frac{{{k^2}}}{{{w^2}}} - {\varepsilon _{22}}{\mu _{33}}} \right){E_2} = 0, \label{NLM75a}
\end{equation}  
and
\begin{equation}
\left( {\frac{{{k^2}}}{{{w^2}}} - {\varepsilon _{33}}{\mu _{22}}} \right){E_3} = 0. \label{NLM75b}
\end{equation} 
Here, it is worth to remark that the equations above, (\ref{NLM75a}) and (\ref{NLM75b}), were obtained in the limit ${{\bf B}_0} \gg {{\bf B}_p}$ and $\lambda  \gg |{{\bf B}_0}|$.

As a consequence, we have two different situations: First, if ${\bf E}\ \bot \ {\bf B}_0$ (perpendicular polarization), from (\ref{NLM75b}) $E_3=0$, and from (\ref{NLM75a}) we get $\frac{{{k^2}}}{{{w^2}}} = {\varepsilon _{22}}{\mu _{33}}$. Hence we see that the dispersion relation of the photon takes the form

\begin{equation}
{n_ \bot } = \sqrt {1 + i\frac{\lambda }{{2\sqrt {{\bf B}_0^2} }}}. \label{NLM80}
\end{equation}
Second, if ${\bf E}\ || \ {\bf B}_0$ (parallel polarization), from (\ref{NLM75a}) $E_2=0$, and from (\ref{NLM75b}) we get $\frac{{{k^2}}}{{{w^2}}} = {\varepsilon _{33}}{\mu _{22}}$. In this case, the corresponding dispersion relation becomes
\begin{equation}
{n_\parallel } = \sqrt {1  }.  \label{NLM85}
\end{equation}
We then easily verify that there are two optical features for the model under consideration. First, the electromagnetic waves with different polarizations have different velocities or, more precisely, the vacuum birefringence phenomenon is present. The second point is related to  the existence of an imaginary part of the index of  refraction which gives rises to vacuum dichroism. As is well known, this refers to the absorption of photons in a vacuum depending on photon polarization.

\section{Interaction energy}

With these considerations in mind, we shall now examine the interaction energy between static point-like sources for the model under study. To this end, we will calculate the expectation value of the energy operator $ H$ in the physical state $ |\Phi\rangle$, along the lines of Refs. \cite{Nonlinear,Logarithmic,Nonlinear2}. Here it is worth emphasizing that, consistent with the approximation of the previous section ($\lambda  \gg \sqrt { - {\cal F}}$), the initial point of our analysis is the Lagrangian density:
\begin{equation}
{\cal L} = ( \frac{1}{{2}})\left( { - {\cal F}} \right) - \frac{\lambda }{{\sqrt 2  }}\sqrt { - {\cal F}}.  \label{Inter05}
\end{equation}

As we have indicated in \cite{Nonlinear,Logarithmic,Nonlinear2}, to handle the exponent $  
{\raise0.5ex\hbox{$\scriptstyle 1$}
\kern-0.1em/\kern-0.15em
\lower0.25ex\hbox{$\scriptstyle 2$}}$ in expression (\ref{Inter05}), we incorporate an auxiliary field $v$ such that its equation of motion gives back the original theory. Therefore the corresponding Lagrangian density takes the form
\begin{equation}
{\cal L} =  - \frac{1}{4}\left( { \frac{1 }{{2{}}} - \frac{{2\lambda }}{{\sqrt 2  }}v} \right){F_{\mu \nu }}{F^{\mu \nu }} 
- \frac{\lambda }{{8\sqrt 2  }}\frac{1}{v}. \label{Inter10}
\end{equation}
With the redefinition $\frac{1}{V} =  \frac{1 }{{2{}}} - \frac{{2\lambda }}{{\sqrt 2}}v$, equation (\ref{Inter10}) becomes
\begin{equation}
{\cal L} =  - \frac{1}{4}\frac{1}{V}{F_{\mu \nu }}{F^{\mu \nu }} - \frac{1}{8}\frac{V}{{\left[ {\frac{V}{2} - 1} \right]}}. \label{Inter15}
\end{equation}

It is worthwhile sketching at this point the canonical quantization of this theory from the Hamiltonian analysis point of view. It may now easily be verified that the canonical momenta are $    
{\Pi ^\mu } =  - \frac{1}{V}{F^{0\mu }}$, so one immediately identifies the two primary constraints 
$\Pi ^0  = 0$ and $p \equiv \frac{{\partial L}}{{\partial \dot v}} = 0$. Furthermore, the momenta are ${\Pi _i} = \frac{1}{V}{E_i}$. Here ${E_i} = {F_{i0}}$. In such a case, the canonical Hamiltonian reduces to
\begin{eqnarray}
{H_C} &=& \int {{d^3}x} \left\{ {{\Pi _i}{\partial ^i}{A_0} + \frac{V}{2}{{\bf \Pi} ^2} + \frac{1}{{2V}}{{\bf B}^2}} \right\} \nonumber\\
 &+& \frac{{{1}}}{{8{}}}\int {{d^3}x} \frac{V}{{\left[ {{\raise0.5ex\hbox{$\scriptstyle 1 $}
\kern-0.1em/\kern-0.15em
\lower0.25ex\hbox{$\scriptstyle {2{}}$}}V - 1} \right]}}. \label{Inter20}
\end{eqnarray}

Next, we also notice that by requiring the primary constraint $\Pi^{0}$ to be preserved in time, one obtains the secondary constraint $\Gamma _1  = \partial _i \Pi ^i  = 0$. Similarly for the constraint $p$, we get the auxiliary field $v$ as 
\begin{equation}
V = 2\left( {1 + \frac{\lambda }{2}\frac{1}{{\sqrt {{{\bf \Pi} ^2}} }}} \right), \label{Inter25}
\end{equation}
which will be used to eliminate $V$. We observe that to get this last expression we have ignored the magnetic field in equation (\ref{Inter20}), because it add nothing to the static potential calculation, as we will show below. According to usual procedure, the corresponding total Hamiltonian that generates the time evolution of the dynamical variables is $  
H = H_C  + \int {d^2 x} \left( {u_0(x) \Pi_0(x)  + u_1(x) \Gamma _1(x) } \right)$, where $u_o(x)$ and $u_1(x)$ are the Lagrange multiplier utilized to implement the constraints. It is a simple matter to verify that $\dot A_0 \left( x \right) = \left[ {A_0 \left( x \right),H} \right] = u_0 \left( x \right)$, which is an arbitrary function. Since $\Pi^0=0$ always, neither $A^0$ nor $\Pi^0$ are of interest in describing the system and may be discarded from the theory. Hence, we can write
\begin{eqnarray}
{H} &=& \int {{d^3}x} \left\{ {w(x){\partial ^i}{\Pi _i} + \frac{V}{2}{{\bf \Pi} ^2} } \right\} \nonumber\\
&+& \frac{{{1}}}{{8{}}}\int {{d^3}x} \frac{V}{{\left[ {{\raise0.5ex\hbox{$\scriptstyle 1 $}
\kern-0.1em/\kern-0.15em
\lower0.25ex\hbox{$\scriptstyle {2{}}$}}V - 1} \right]}}, \label{Inter30}
\end{eqnarray}
where $w(x) = u_1 (x) - A_0 (x)$ and $V$ is given by (\ref{Inter25}).

We can at this stage impose a gauge condition, so that in conjunction with the constraint ${\Pi ^0} = 0$, it is rendered into a second class set. A particularly convenient choice is
\begin{equation}
\Gamma _2 \left( x \right) \equiv \int\limits_{C_{\xi x} } {dz^\nu
} A_\nu \left( z \right) \equiv \int\limits_0^1 {d\lambda x^i }
A_i \left( {\lambda x} \right) = 0. \label{Inter35}
\end{equation}
where  $\lambda$ $(0\leq \lambda\leq1)$ is the parameter describing
the spacelike straight path $ z^i = \xi ^i  + \lambda \left( {x -\xi } \right)^i $, and $ \xi $ is a fixed point (reference point). We also recall that there is no essential loss of generality if we restrict our considerations to $ \xi ^i=0 $. Hence the only nontrivial Dirac bracket for the canonical variables is given by
\begin{eqnarray}
\left\{ {A_i \left( x \right),\Pi ^j \left( y \right)} \right\}^ * &=& \delta _i^j \delta ^{\left( 3 \right)} \left( {x - y} \right) \nonumber\\
&-&\partial _i^x \int\limits_0^1 {d\lambda x^i } \delta ^{\left( 3\right)} \left( {\lambda x - y} \right). \label{Inter40}
\end{eqnarray}

We now proceed to compute the interaction energy for the model under consideration. As mentioned above, to do that we need to compute the expectation value of the energy operator $H$ in the physical state $\left| \Phi  \right\rangle$. Following Dirac \cite{Dirac}, we write the physical state $\left| \Phi  \right\rangle$ as
\begin{equation}
\left| \Phi  \right\rangle  \equiv \left| {\bar \Psi ({\bf y})\Psi ({{\bf y}^ \prime })} \right\rangle  = \bar \psi ({\bf y})\exp (ie\int_{{{\bf y}^ \prime }}^{\bf y} {d{z^i}{A_i}(z)} )\psi ({{\bf y}^ \prime })\left| 0 \right\rangle,                              
\label{Inter45}
\end{equation}
where $\left| 0 \right\rangle$ is the physical vacuum state and the
line integral appearing in the above expression is along a spacelike
path starting at ${\bf y}\prime$ and ending at $\bf y$, on a fixed
time slice. The above expression clearly shows that, each of the states $(\left| \Phi  \right\rangle)$, 
represents a fermion-antifermion pair surrounded by a cloud of gauge fields to maintain gauge invariance.

Taking the above Hamiltonian structure into account, we see that
\begin{eqnarray}
\Pi _i \left( x \right)\left| {\overline \Psi \left( \bf{y }\right)\Psi\left( {\bf{y}^ \prime } \right)} \right\rangle &=& \overline \Psi \left( \bf{y }\right)\Psi \left( {\bf{y}^ \prime } \right)\Pi _i \left( x\right)\left| 0 \right\rangle \nonumber\\
&+&\int_{\bf y}^{{\bf y}\prime} {d{z_i}{\delta ^{\left( 3 \right)}}\left( {{\bf z} - {\bf x}} \right)\left| \Phi  \right\rangle }.
\label{Inter50}
\end{eqnarray}

As a consequence of this, by employing  (\ref{Inter50}), (\ref{Inter30}) and (\ref{Inter25}), the interaction energy takes the form
\begin{equation}
{\left\langle H \right\rangle _\Phi } = {\left\langle H \right\rangle _0} + {V_1} + {V_2}, \label{Inter55}
\end{equation}
where ${\left\langle H \right\rangle _0} = \left\langle 0 \right|H\left| 0 \right\rangle$. The $V_1$ and $V_2$ are given by
\begin{equation}
{V_1} = \int {{d^3}x\left\langle \Phi  \right|} {{\bf \Pi} ^2}\left| \Phi  \right\rangle, \label{Inter60}
\end{equation}
and
\begin{equation}
{V_2} = {\lambda} \int {{d^3}x\left\langle  \Phi  \right|\sqrt {{{\bf \Pi} ^2}} } \left| \Phi  \right\rangle. \label{Inter65} 
\end{equation}
At this point we should mention that the reason why we eliminated from the Hamiltonian the magnetic field now becomes clear, that is, the commutator for the magnetic field is zero.

Following our earlier procedure \cite{Gaete,GaeteGueSpa}, the static potential turns out to be
\begin{equation}
V =  - \frac{{{e^2}}}{{4\pi }} \frac{2}{r} + e {\lambda} r, \label{Inter70}
\end{equation}
after subtracting a self-energy term.

Before concluding this subsection it is constructive to briefly examine an alternative derivation of our previous result, which permits us to check the internal consistency of our procedure. In order to illustrate the discussion, we begin by recalling that
\begin{equation}
V \equiv e\left( {{\cal A}_0 \left( {\bf 0} \right) - {\cal A}_0 \left( {\bf L} \right)} \right), \label{BIL100}
\end{equation}
where the physical scalar potential is given by
\begin{equation}
{\cal A}_0 (t,{\bf r}) = \int_0^1 {d\lambda } r^i E_i (t,\lambda
{\bf r}). \label{Inter75}
\end{equation}
This equation follows from the vector gauge-invariant field expression
\begin{equation}
{\cal A}_\mu  (x) \equiv A_\mu  \left( x \right) + \partial _\mu  \left( { - \int_\xi ^x {dz^\mu  A_\mu  \left( z \right)} } \right), \label{Inter80}
\end{equation}
where the line integral is along a spacelike path from the point $\xi$ to $x$, on a fixed slice time. It should again be stressed here that the gauge-invariant variables (\ref{Inter80}) commute with the sole first constraint (Gauss law), showing in this way that these fields are physical variables. 

It should be noted that Gauss' law for the present theory reads
\begin{equation}
{\partial _i}{\Pi ^i} = {J^0},  \label{Inter85}
\end{equation}
where ${E^i} =V{\Pi ^i} $ and $V$ is given by equation (\ref{Inter25}). Note that we have included the external current $J^0$ to represent the presence of external charges. In such a case, for ${J^0}({\bf r}) = e{\delta ^{\left( 3 \right)}}\left( {\bf r} \right)$, the electric field reduces to  
\begin{equation}
{\bf E} = \frac{{\left( {2e} \right)}}{{4\pi {r^2}}}\left( {1 + \frac{{2\pi \lambda }}{e}{r^2}} \right)\hat r. \label{Inter90}
\end{equation}

Using (\ref{Inter90}), we can express (\ref{Inter75}) as
\begin{equation}
{{\cal A}_0}(t,{\bf r}) =  - 2 \int_0^r {dz\left( {\frac{e}{{4\pi {z^2}}} + \frac{\lambda }{2}} \right)}, \label{Inter95}
\end{equation}
We can, therefore, write
\begin{equation}
{{\cal A}_0}(t,{\bf r}) = - \frac{e}{{4\pi }}\frac{2}{r} - \lambda r. \label{Inter100}
\end{equation}

Accordingly, by employing Eq. (\ref{Inter100}), finally we end up with the potential for a pair of static point-like opposite charges located at $\bf 0$ and $\bf L$, 
\begin{equation}
V =  - \frac{{{e^2}}}{{4\pi }} \frac{2}{r} + e {\lambda} r,  \label{Inter105}
\end{equation}
after subtracting a self-energy term.\\

\section{Final Remarks}

In summary, we have considered a new nonlinear electrodynamics. It was shown that in this new electrodynamics the phenomenon of birefringence and dichroism take place in the presence of external magnetic fields. Subsequently we have studied the interaction energy. To do this, once again we have exploited a key aspect for understanding the physical contents of gauge theories, that is, the correct identification of field degrees of freedom with observable quantities.
Interestingly enough, our analysis reveals that the static potential profile contains a linear potential leading to the confinement of static charges. It remains to be worked out how to connect our $\lambda$-parameter to the recent measurement of light-by-light scattering \cite{Atlas} and to the PVLAS experiment of light's non-linearity effects. We shall be reporting on that in a forthcoming work.

\section{ACKNOWLEDGMENTS}
One of us (P. G.) wishes to thank the Field Theory Group of the COSMO/CBPF for the hospitality and the PCI-BEV/MCTIC support. P. G. was partially supported by Proyecto Basal FB 0821.

\end{document}